\documentclass[twocolumn,showpacs,preprintnumbers,amsmath,amssymb,groupedaddrss,superscriptaddress]{revtex4}
\usepackage{graphicx}
\usepackage{fancybox}
\usepackage{dcolumn}
\usepackage{epsfig}
\usepackage[dvips]{color}
\usepackage{bm}

\allowdisplaybreaks[4]

\begin{document}
\title{
Vanishing dimension five proton decay operators 
in the $SU(5)$ SUSY GUT
}

\author{Naoyuki Haba}
%\email{haba@ph.tum.de}
\affiliation{Physik Department, Technische Universit\"at M\"unchen, 
James-Franck-Strasse 85748 Garching, Germany}
\affiliation{Institute of Theoretical Physics, University of
Tokushima, 
770-8502 Tokushima, Japan}

\author{Toshihiko Ota}
%\email{toshi@ph.tum.de}
\affiliation{Physik Department, Technische Universit\"at M\"unchen, 
James-Franck-Strasse 85748 Garching, Germany}

\preprint{TUM-HEP-642/06}

\date{\today}


\begin{abstract}
We propose a natural framework of 
 $SU(5)$ supersymmetric grand unified 
 theory with the minimal particle contents, 
 which does not contain dimension five proton
 decay operators. 
The suitable fermion mass hierarchy can be reproduced
 by higher dimensional operators of
 an adjoint Higgs field which breaks
 $SU(5)$ gauge symmetry. 
\end{abstract}

\maketitle

%%%%%%%%%%%%%%%%%%%%%%%%%%%%%%%%%%%%%%%%%%%%%%%%%%%%%%%%%%%%%%%%%%%%%%
%\section{Introduction}

Supersymmetric (SUSY) grand unified theories (GUTs)\cite{SUSYGUT}
have been regarded as the most agreeable candidates  
beyond the standard model for a long time,
because they can realize the gauge coupling unification 
as well as show a natural solution of
the hierarchy problem. 
They can also explain the electroweak symmetry breaking by the so-called 
 the radiative breaking scenario\cite{IKKT}.
The proton decay is a crucial prediction of 
 GUTs\cite{pdecayinSUSYGUT},
 which has not been 
 observed in the experiments yet\cite{currentlimit}.
From the current experimental bound on the
 proton decay, 
 it was claimed that the minimal $SU(5)$ SUSY GUT 
 has already been excluded\cite{GotoNihei,MurayamaPierce}. 
However, as pointed out in e.g. Refs.\cite{examples,consistentmodel},
 this claim should only be applied 
 to the minimal scenario 
 which is unrealistic in the sense that it cannot
 reproduce the correct mass spectrum of down type quarks 
 and charged leptons.
Notice that we must analyze 
 the Yukawa interactions with the coloured Higgs fields 
 carefully. 
In this letter we show the natural model 
 which does not contain dimension five proton decay operators 
 in the $SU(5)$ SUSY GUT framework 
 with the minimal particle 
 content.
We must introduce the GUT scale ($M_{\rm GUT}$) where 
 the $SU(5)$ gauge symmetry is broken 
 by the vacuum expectation value (VEV) of 
 an adjoint representation (${\bf 24}$) 
 Higgs field $\Sigma$.
Since 
 the model at the GUT scale
 is the effective theory of the fundamental one 
 which realized at the Planck scale,
 it should contain the higher dimensional operators which are
 suppressed by the Planck 
 scale $(M_{\rm GUT}/M_{\rm Pl})^n$ ($n:$ positive integer)
 \cite{Nath,higherdimop,consistentmodel}.
These terms can be  
 the origin of a 
 fermion mass hierarchy of three generations as well as 
 the one between top and bottom quarks. 
The realistic mass spectrum of the down type quarks and
 the charged leptons can be reproduced
 by these terms.
We take the following setup;
 \begin{enumerate}
 \item Only the top Yukawa coupling exists at the tree level.
       The other Yukawa couplings are 
       induced by the $n${\it th order} higher dimensional terms,
       $(M_{\rm GUT}/M_{\rm Pl})^n$.
       
 \item The Yukawa couplings of the bottom quark and 
       the tau lepton are induced by 
        {\it the first order} terms $(M_{\rm GUT}/M_{\rm Pl})$, 
       the mass of the strange quark and the muon are reproduced by
       {\it the second order} 
       terms $(M_{\rm GUT}/M_{\rm Pl})^2$, 
       and 
       the down quark and the electron 
       masses are provided by {\it the third order} 
       terms $(M_{\rm GUT}/M_{\rm Pl})^3$. 
       The masses of the up type quarks might be appropriately 
       reproduced in a similar way.
       
 \item We should pay attention 
       that some of the terms 
       $(\langle \Sigma \rangle / M_{\rm Pl})^{n}$ 
       are regarded as 
       $(n\hspace{-1mm}-\hspace{-1mm}1)${\it th order}
       terms effectively due to  
       their coefficients. 
       
  \item 
       The couplings of the operators associated with the proton decay
       process completely vanish.
 \end{enumerate}
%As will be shown below,
% some higher dimensional operators have 
% large pre-factors (coefficients), which 
% will be regarded as $(n\hspace{-1mm}-\hspace{-1mm}1)$$th$ $order$
% operators.  
%We will show the Yukawa couplings of the coloured Higgs triplets 
% which has no dimension five proton decay operator,
% as keeping the realistic mass spectrum for the 
% matter fermions. 
%It is necesary to care about {\it the order} of 
%the higher dimensional terms.
%Depending on the gauge group structure of  
%the interactions between the matter fields and 
%the $SU(5)$ breaking Higgs field,
%some of the terms take larger 
%
%We should take into account them when 
% they are so large as to compensate  
% the suppression factor $M_{\rm GUT}/M_{\rm Pl}$.
As we will show,
the two requirements --- the realistic Yukawa
 couplings of quarks and leptons 
 and no dimension five proton decay operators --- 
 almost determine the couplings for the higher
 dimensional operators.

%%%%%%%%%%%%%%%%%%%%%%%%%%%%%%%%%%%%%%%%%%%%%%%%%%%%%%%%%%%%%%%%%%%%%%
%\section{Model}

The superpotential for the Yukawa sector is represented 
 as the series expansion according to the power of 
 adjoint Higgs fields, such as
%Our basic concepts are as follows; 
%only the top quark mass is induced from the zeroth 
%order, 
%the bottom quark and tau lepton Yukawa couplings are 
%induced from the first order,
%and the second (first) generation masses are from the second (third) 
%order.
%As shown below,
%some of the terms in the fourth order of $M_{\rm GUT}/M_{\rm Pl}$
%contribute as large as third oder terms
%because of their large coefficients.
%Therefore, we take them up to the fourth order of $M_{\rm GUT}/M_{\rm Pl}$.
\begin{align}
W_{\text{Yukawa}}
=&
W_{0}
+ W_{1} 
+ W_{2}
+ W_{3}
+ W_{4}
+ \cdots.
\end{align}
The zeroth order part $W_{0}$
is the same as that of the minimal $SU(5)$ model,
\begin{align}
W_{0} =&
 \frac{1}{4}
 \epsilon_{abcde}
 Y_{1}^{ij}
 {\bf 10}^{ab}_{i} {\bf 10}^{cd}_{j} H^{e}
 +
 \sqrt{2}
 Y_{2}^{ij} 
 \bar{H}_{a}
 {\bf 10}^{ab}_{i}
 {\bf 5}^{*}_{j b},
\label{2}
\end{align}
where $i,j$ are the indices for generations and 
$a,b,...$ are the ones for the $SU(5)$ indices. %fundamental representation.
The chiral superfield {\bf 10} contains 
the right-handed up-type quark $u_{R}^{\rm c}$,
the left-handed quark doublet $Q$,
and the right-handed charged lepton $e_{R}^{\rm c}$.
The right-handed down-type quark $d_{R}^{\rm c}$
and the lepton doublet $L$ belong to the superfield ${\bf 5}^{*}$.  
The Higgs fields $H$ and $\bar{H}$ 
(${\bf 5} $ and  ${\bf 5}^{*}$)
 include the coloured Higgs triplets ($H_{C}$, $\bar{H}_C$) and 
 the Higgs doublets, respectively. 
As shown in our setup, 
 $Y_2=0$ is assumed in Eq.\eqref{2}, 
 which 
 is the origin of the hierarchy 
 between the top and the bottom masses. 
The {\it first order} part $W_{1}$
is expressed as\cite{consistentmodel} 
\begin{align}
&W_{1} =
\frac{\epsilon_{abcde}}{4}
%\epsilon_{abcde}
\left(
 f_{1}^{ij}
 {\bf 10}^{ab}_{i} {\bf 10}^{cd}_{j}
 \frac{\Sigma^{e}_{f}}{M_{\rm Pl}} 
 H^{f}
 +
 f_{2}^{ij}
 {\bf 10}^{ab}_{i} {\bf 10}^{cf}_{j} 
  H^{d}
 \frac{\Sigma^{e}_{f}}{M_{\rm Pl}} 
 \right) \nonumber \\
%%%%%
%& \hspace{2cm}
% \left.
% +f_{0}^{ij}
% {\bf 10}^{ab}_{i} {\bf 10}^{cd}_{j} H^{e}
% \frac{\phi}{M_{\rm Pl}}
% \right)  \nonumber \\
%%%%%
&\hspace{0.2cm}+
 \sqrt{2}
 \left(
 h_{1}^{ij} \bar{H}_{a}
 \frac{\Sigma_{b}^{a}}{M_{\rm Pl}}
 {\bf 10}^{bc}_{i}
 {\bf 5}^{*}_{jc} 
 +
 h_{2}^{ij} \bar{H}_{a}
 {\bf 10}^{ab}_{i}
 \frac{\Sigma_{b}^{c}}{M_{\rm Pl}}
 {\bf 5}^{*}_{jc} \right) ,% \nonumber \\
%%%%%
% &\left.\hspace{1cm}
% + h_{0}^{ij} 
% \bar{H}_{a}
% {\bf 10}^{ab}_{i}
% {\bf 5}^{*}_{j b}
% \frac{\phi}{M_{\rm Pl}}
% \right),
\label{eq:W-1}
\end{align}
where $\Sigma$ takes VEV of 
 $\langle \Sigma \rangle = {\rm diag} (2,2,2,-3,-3) \sigma$ 
 which 
 breaks the $SU(5)$ gauge group into 
 $SU(3)_{\rm c} \times SU(2)_{L} \times U(1)_{Y}$. 
The value of $\sigma$ is the scale of $M_{\rm GUT}$.  
For the {\it second} and {\it third order} superpotential, 
% of $\Sigma/M_{\rm Pl}$,
 we only show the down-type quark and the charged lepton sector.
They are represented as\cite{Nath}
\begin{align}
&W_{2}
=
 \sqrt{2}
 \Biggl(
 h_{3}^{ij}
 \bar{H}_{a} {\bf 10}^{ab}_{i} {\bf 5}^{*}_{jb}
 \frac{(\Sigma \Sigma)}{M_{\rm Pl}^{2}}
% \frac{\Sigma_{d}^{c}}{M_{\rm Pl}} \nonumber \\
%&  \hspace{1cm}
 +
 h_{4}^{ij}
 \bar{H}_{a} 
 \frac{(\Sigma \Sigma)^{a}_{b}}{M_{\rm Pl}^{2}}
% \frac{\Sigma_{b}^{d}}{M_{\rm Pl}}
 {\bf 10}^{bc}_{i} {\bf 5}^{*}_{jc} \nonumber \\
& +
 h_{5}^{ij}
 \bar{H}_{a} 
 {\bf 10}^{ab}_{i} 
 \frac{(\Sigma \Sigma)_{b}^{c}}{M_{\rm Pl}^{2}}
% \frac{\Sigma_{b}^{d}}{M_{\rm Pl}}
 {\bf 5}^{*}_{jc} %\nonumber \\
%&  \hspace{1cm}
+
 h_{6}^{ij}
 \bar{H}_{a} 
 \frac{\Sigma_{b}^{a}}{M_{\rm Pl}}
 {\bf 10}^{bc}_{i} 
 \frac{\Sigma_{c}^{d}}{M_{\rm Pl}}
  {\bf 5}^{*}_{jd} %\nonumber \\
%%%%
% & \hspace{0.5cm}
% + 
% h_{\phi_{1}}^{ij} \bar{H}_{a}
% \frac{\Sigma_{b}^{a} \phi}{M_{\rm Pl}^{2}}
% {\bf 10}^{bc}_{i}
% {\bf 5}^{*}_{jc} 
% +
% h_{\phi_{2}}^{ij} \bar{H}_{a}
% {\bf 10}^{ab}_{i}
% \frac{\Sigma_{b}^{c} \phi }{M_{\rm Pl}^{2}}
% {\bf 5}^{*}_{jc} \nonumber \\
%%%%%
% &\hspace{0.5cm}
% + h_{\phi_{0}}^{ij} 
% \bar{H}_{a}
% {\bf 10}^{ab}_{i}
% {\bf 5}^{*}_{j b}
% \frac{ \phi^{2}}{M_{\rm Pl}^{2}}
 \Biggr), 
\label{eq:W-2}\\
%%%%%
 &W_{3} =
 \sqrt{2}
 \Biggl(
 h_{7}^{ij} 
 \bar{H}_{a} 
 \frac{\Sigma_{b}^{a}}{M_{\rm Pl}}
 {\bf 10}^{bc}_{i} {\bf 5}^{*}_{jc}
 \frac{(\Sigma \Sigma)}{M_{\rm Pl}^{2}} 
 \nonumber \\
& 
 +
 h_{8}^{ij} 
 \bar{H}_{a} 
 {\bf 10}^{ab}_{i}
 \frac{\Sigma_{b}^{c}}{M_{\rm Pl}}
 {\bf 5}^{*}_{jc}
 \frac{(\Sigma \Sigma)}{M_{\rm Pl}^{2}} 
 +
 h_{9}^{ij} 
 \bar{H}_{a} 
 {\bf 10}^{ab}_{i}
 {\bf 5}^{*}_{jb}
 \frac{(\Sigma \Sigma \Sigma)}{M_{\rm Pl}^{3}} 
 \nonumber \\
%%%%%
& +
 h_{10}^{ij} 
 \bar{H}_{a} 
 \frac{(\Sigma \Sigma \Sigma)^{a}_{b}}{M_{\rm Pl}^{3}}
 {\bf 10}^{bc}_{i}
 {\bf 5}^{*}_{jc}
 +
 h_{11}^{ij} 
 \bar{H}_{a} 
 {\bf 10}^{ab}_{i}
 \frac{(\Sigma \Sigma \Sigma)_{b}^{c}}{M_{\rm Pl}^{3}}
 {\bf 5}^{*}_{jc}
 \nonumber \\
%%%%%
& +
 h_{12}^{ij}
 \bar{H}_{a} 
 \frac{(\Sigma \Sigma)_{b}^{a}}{M_{\rm Pl}^{2}}
 {\bf 10}^{bc}_{i}
 \frac{\Sigma_{c}^{d}}{M_{\rm Pl}}
 {\bf 5}^{*}_{jd} 
 +
  h_{13}^{ij}
 \bar{H}_{a} 
 \frac{\Sigma_{b}^{a}}{M_{\rm Pl}}
 {\bf 10}^{bc}_{i}
 \frac{(\Sigma \Sigma)_{c}^{d}}{M_{\rm Pl}^{2}}
 {\bf 5}^{*}_{jd} 
 \Biggr),
% \Biggl(
% h_{\phi_{3}}^{ij}
% \bar{H}_{a} {\bf 10}^{ab}_{i} {\bf 5}^{*}_{jb}
% \frac{\Sigma_{c}^{d} \Sigma_{d}^{c} \phi }{M_{\rm Pl}^{3}}
%% \frac{\Sigma_{d}^{c}}{M_{\rm Pl}} \nonumber \\
%%&  \hspace{1cm}
% +
% h_{\phi_{4}}^{ij}
% \bar{H}_{a} 
% \frac{\Sigma_{d}^{a} \Sigma_{b}^{d} \phi}{M_{\rm Pl}^{3}}
%% \frac{\Sigma_{b}^{d}}{M_{\rm Pl}}
% {\bf 10}^{bc}_{i} {\bf 5}^{*}_{jc} \nonumber \\
%& +
% h_{\phi_{5}}^{ij}
% \bar{H}_{a} 
% {\bf 10}^{ab}_{i} 
% \frac{\Sigma_{d}^{c} \Sigma_{b}^{d} \phi }{M_{\rm Pl}^{3}}
%% \frac{\Sigma_{b}^{d}}{M_{\rm Pl}}
% {\bf 5}^{*}_{jc} %\nonumber \\
%%&  \hspace{1cm}
%+
% h_{\phi_{6}}^{ij}
% \bar{H}_{a} 
%% \frac{\Sigma_{b}^{a}}{M_{\rm Pl}}
% {\bf 10}^{bc}_{i} 
% \frac{\Sigma_{b}^{a} \Sigma_{c}^{d} \phi}{M_{\rm Pl}^{3}}
%  {\bf 5}^{*}_{jd} \\
%%%%%%
%&+
% h_{\phi_{1}^{2}}^{ij} \bar{H}_{a}
% \frac{\Sigma_{b}^{a} \phi^{2}}{M_{\rm Pl}^{2}}
% {\bf 10}^{bc}_{i}
% {\bf 5}^{*}_{jc} 
% +
% h_{\phi_{2}^{2}}^{ij} \bar{H}_{a}
% {\bf 10}^{ab}_{i}
% \frac{\Sigma_{b}^{c} \phi^{2} }{M_{\rm Pl}^{2}}
% {\bf 5}^{*}_{jc} \nonumber \\
%%%%%%
% &
% + h_{\phi_{0}^{2}}^{ij} 
% \bar{H}_{a}
% {\bf 10}^{ab}_{i}
% {\bf 5}^{*}_{j b}
% \frac{ \phi^{3}}{M_{\rm Pl}^{2}}
%+ \cdots\Biggr) .
\label{eq:W-3}
\end{align} 
where $(\Sigma\cdots)$ ($(\Sigma \cdots)_{b}^{a}$) denotes 
 a singlet (an adjoint) by 
 contracting the $SU(5)$ indices of $\Sigma \cdots$.  
The {\it fourth order} superpotential
 suggests 
\begin{align}
&W_{4} =
 \sqrt{2}\Biggl(
 h_{14}^{ij}
 \bar{H}_{a} {\bf 10}^{ab} {\bf 5}^{*}_{b}
 \frac{(\Sigma \Sigma)(\Sigma \Sigma)}{M_{\rm Pl}^{4}}
 \nonumber \\
%%%%%
& +
 h_{15}^{ij}
 \bar{H}_{a} {\bf 10}^{ab} {\bf 5}^{*}_{b}
 \frac{(\Sigma \Sigma \Sigma \Sigma)}{M_{\rm Pl}^{4}}
+
 h_{16}^{ij}
 \bar{H}_{a} 
 \frac{\Sigma_{b}^{a}}{M_{\rm Pl}}
 {\bf 10}^{bc} {\bf 5}^{*}_{c}
 \frac{(\Sigma\Sigma\Sigma)}{M_{\rm Pl}^{3}}
 \nonumber \\
%%%%%
 &
+
 h_{17}^{ij}
  \bar{H}_{a} 
 {\bf 10}^{ab}
 \frac{\Sigma_{b}^{c}}{M_{\rm Pl}}
 {\bf 5}^{*}_{c}
 \frac{(\Sigma\Sigma\Sigma)}{M_{\rm Pl}^{3}}
 +
 h_{18}^{ij}
 \bar{H}_{a} 
 \frac{(\Sigma\Sigma)_{b}^{a}}{M_{\rm Pl}^{2}}
 {\bf 10}^{bc} {\bf 5}^{*}_{c}
 \frac{(\Sigma\Sigma)}{M_{\rm Pl}^{2}}
 \nonumber \\
%%%%%
& +
 h_{19}^{ij}
  \bar{H}_{a} 
 {\bf 10}^{ab}
 \frac{(\Sigma\Sigma)_{b}^{c}}{M_{\rm Pl}^{2}}
 {\bf 5}^{*}_{c}
 \frac{(\Sigma\Sigma)}{M_{\rm Pl}^{2}}
 +
  h_{20}^{ij}
 \bar{H}_{a} 
 \frac{\Sigma_{b}^{a}}{M_{\rm Pl}}
 {\bf 10}^{bc} 
 \frac{\Sigma_{c}^{d}}{M_{\rm Pl}}
 {\bf 5}^{*}_{d}
 \frac{(\Sigma\Sigma)}{M_{\rm Pl}^{2}}
 \nonumber \\
%%%%%
&  +
  h_{21}^{ij}
 \bar{H}_{a} 
 \frac{(\Sigma\Sigma\Sigma\Sigma)_{b}^{a}}{M_{\rm Pl}^{4}}
 {\bf 10}^{bc} 
  {\bf 5}^{*}_{c}
+
  h_{22}^{ij}
 \bar{H}_{a} 
 {\bf 10}^{ab} 
 \frac{(\Sigma\Sigma\Sigma\Sigma)_{b}^{c}}{M_{\rm Pl}^{4}}
  {\bf 5}^{*}_{c} 
\nonumber \\
%%%%%
&+
  h_{23}^{ij}
 \bar{H}_{a} 
 \frac{(\Sigma\Sigma)_{b}^{a}}{M_{\rm Pl}^{2}}
 {\bf 10}^{bc} 
 \frac{(\Sigma\Sigma)_{c}^{d}}{M_{\rm Pl}^{2}}
  {\bf 5}^{*}_{d}
+
  h_{24}^{ij}
 \bar{H}_{a} 
 \frac{(\Sigma\Sigma\Sigma)_{b}^{a}}{M_{\rm Pl}^{3}}
 {\bf 10}^{bc} 
 \frac{\Sigma_{c}^{d}}{M_{\rm Pl}}
  {\bf 5}^{*}_{d} \nonumber \\
%%%%%
&+
  h_{25}^{ij}
 \bar{H}_{a} 
 \frac{\Sigma_{b}^{a}}{M_{\rm Pl}}
 {\bf 10}^{bc} 
 \frac{(\Sigma\Sigma\Sigma)_{c}^{d}}{M_{\rm Pl}^{3}}
  {\bf 5}^{*}_{d}
\Biggr).
\label{eq:W-4}
\end{align} 
Each matrix element $h^{ij}$
 is assumed to have an $\mathcal{O}(1)$ coefficient,
 and the mass hierarchy is produced by 
 the suppression factors 
 $\sigma/M_{\rm Pl} \equiv 1/a$\footnote{%
The authors of Ref.[8] have also 
 introduced
 higher dimensional terms but up to the first 
 order of $1/a$. 
In this case, there is an inevitable
 relation between the Yukawa matrices as
 $Y_{ql} - Y_{ud} = Y_{e} - Y_{d} \neq 0$, where 
 the dimension five operators cannot 
 completely vanish 
 and the suppression factor $1/a$ has nothing to do with
 the fermion mass hierarchy. 
On the other hand, 
 the author of Ref.[9] has tried to reproduce 
 the so called Georgi-Jarlskog texture
[See H. Georgi and C. Jarlskog, Phys. Lett. {\bf B86} (1979) 297] 
by introducing the terms up to the third order of $1/a$.
}.

Decomposing the superpotential 
Eqs.\eqref{eq:W-1}-\eqref{eq:W-4}
into its component fields, 
we obtain the Yukawa couplings 
of down-type quarks and charged leptons as
\begin{align}
 Y_{d} =& 
 Y' 
 -
 \frac{3}{a} h_{1}'
 +
 \frac{2}{a} h_{2}'
% + 
% \frac{30}{a^{2}} h_{3} 
 +
 \frac{9}{a^{2}} 
 h_{4}'
 +
 \frac{4}{a^{2}} 
 h_{5}'
 -
 \frac{6}{a^{2}} 
 h_{6}' \nonumber \\
%%%%%
&
% -
% \frac{90}{a^{3}} h_{7}
% +
% \frac{60}{a^{3}} h_{8}
% -
% \frac{30}{a^{3}} h_{9}
 -
 \frac{27}{a^{3}} h_{10} 
 +
 \frac{8}{a^{3}} h_{11}
 +
 \frac{18}{a^{3}} h_{12} 
 -
 \frac{12}{a^{3}} h_{13} +\cdots,  \nonumber \\
%& 
%+
% \frac{900}{a^{4}} h_{14} 
% +
% \frac{210}{a^{4}} h_{15}
% +
% \frac{90}{a^{4}} h_{16}
% -
% \frac{60}{a^{4}} h_{17}
% +
% \frac{270}{a^{4}} h_{18} 
% +
% \frac{120}{a^{4}} h_{19} %\nonumber \\
% -
% \frac{180}{a^{4}} h_{20} + \cdots,
% +
% \frac{81}{a^{4}} h_{21}
% +
% \frac{16}{a^{4}} h_{22}
% +
% \frac{36}{a^{4}} h_{23} 
% -
% \frac{54}{a^{4}} h_{24}
% -
% \frac{24}{a^{4}} h_{25},
\label{eq:Yd}\\
%%%%%
 Y_{e} = &
 Y' 
 -
 \frac{3}{a} h_{1}'
 -
 \frac{3}{a} h_{2}'
% +
% \frac{30}{a^{2}} h_{3}
 +
 \frac{9}{a^{2}} h_{4}'
 +
 \frac{9}{a^{2}} h_{5}'
 +
 \frac{9}{a^{2}} h_{6}' \nonumber \\
%%%%%
 &
% -
% \frac{90}{a^{3}} h_{7}
% -
% \frac{90}{a^{3}} h_{8}
% -
% \frac{30}{a^{3}} h_{9}
 -
 \frac{27}{a^{3}} h_{10} 
 -
 \frac{27}{a^{3}} h_{11}
 -
 \frac{27}{a^{3}} h_{12}
 -
 \frac{27}{a^{3}} h_{13} + \cdots, %\nonumber\\
%%%%%
%& 
%+
% \frac{900}{a^{4}} h_{14} 
% +
% \frac{210}{a^{4}} h_{15}
% +
% \frac{90}{a^{4}} h_{16}
% +
% \frac{90}{a^{4}} h_{17}
% +
% \frac{270}{a^{4}} h_{18} 
% +
% \frac{270}{a^{4}} h_{19} %\nonumber \\
% +
% \frac{270}{a^{4}} h_{20} + \cdots,
% +
% \frac{81}{a^{4}} h_{21}
% +
% \frac{81}{a^{4}} h_{22}
% +
% \frac{81}{a^{4}} h_{23} 
% +
% \frac{81}{a^{4}} h_{24}
% +
% \frac{81}{a^{4}} h_{25},
\label{eq:Ye}
\end{align}
where the matrices with a prime symbol are defined as 
\begin{align}
Y' \equiv&
 Y_{2}
 +
 \frac{30}{a^{2}}
 h_{3}
 -
 \frac{30}{a^{3}}
 h_{9}
 +
 \frac{900}{a^{4}}
 h_{14}
 +
 \frac{210}{a^{4}}
 h_{15}, \\
%%%%%
h_{1,2}' \equiv&
 h_{1,2} + \frac{30}{a^{2}} h_{7,8} - \frac{30}{a^{3}} h_{16,17},
\\
%%%%%
h_{4,5,6}' \equiv&
 h_{4,5,6} + \frac{30}{a^{2}} h_{18,19,20}.
\end{align}
On the other hand, 
 the couplings $Y_{ql}$ and $Y_{ud}$ 
 which are associated with the interactions 
 $Q_{i} \epsilon L_{j} \bar{H}_{C}$ and
 $u_{Ri}^{\rm c} d_{Rj}^{\rm c} \bar{H}_{C}$, respectively, 
 are given by 
\begin{align}
 Y_{ql} =& 
 Y' 
 +
 \frac{2}{a} h_{1}'
 -
 \frac{3}{a} h_{2}'
% +
% \frac{30}{a^{2}} h_{3}
 +
 \frac{4}{a^{2}} h_{4}'
 +
 \frac{9}{a^{2}} h_{5}'
 -
 \frac{6}{a^{2}} h_{6}' \nonumber \\
%%%%%
 & 
% +
% \frac{60}{a^{3}} h_{7}
% -
% \frac{90}{a^{3}} h_{8}
% -
% \frac{30}{a^{3}} h_{9}
 +
 \frac{8}{a^{3}} h_{10}
 -
 \frac{27}{a^{3}} h_{11}
 -
 \frac{12}{a^{3}} h_{12}
 +
 \frac{18}{a^{3}} h_{13} + \cdots, %\nonumber\\
%%%%%
%& 
%+
% \frac{900}{a^{4}} h_{14} 
% +
% \frac{210}{a^{4}} h_{15}
% -
% \frac{60}{a^{4}} h_{16}
% +
% \frac{90}{a^{4}} h_{17}
% +
% \frac{120}{a^{4}} h_{18} 
% +
% \frac{270}{a^{4}} h_{19} %\nonumber \\
% -
% \frac{180}{a^{4}} h_{20} + \cdots,
% +
% \frac{16}{a^{4}} h_{21}
% +
% \frac{81}{a^{4}} h_{22}
% +
% \frac{36}{a^{4}} h_{23} 
% -
% \frac{24}{a^{4}} h_{24}
% -
% \frac{54}{a^{4}} h_{25},
\label{eq:Yql}\\
%%%%%
 Y_{ud} =& 
 Y' 
 +
 \frac{2}{a} h_{1}'
 +
 \frac{2}{a} h_{2}'
% +
% \frac{30}{a^{2}} h_{3}
 +
 \frac{4}{a^{2}} h_{4}'
 +
 \frac{4}{a^{2}} h_{5}'
 +
 \frac{4}{a^{2}} h_{6}' \nonumber \\
%%%%%
 & 
% +
% \frac{60}{a^{3}} h_{7}
% +
% \frac{60}{a^{3}} h_{8}
% -
% \frac{30}{a^{3}} h_{9}
 +
 \frac{8}{a^{3}} h_{10} 
 +
 \frac{8}{a^{3}} h_{11}
 +
 \frac{8}{a^{3}} h_{12}
 +
 \frac{8}{a^{3}} h_{13} + \cdots. % \nonumber \\
 %%%%%
%& 
%+
% \frac{900}{a^{4}} h_{14} 
% +
% \frac{210}{a^{4}} h_{15}
% -
% \frac{60}{a^{4}} h_{16}
% -
% \frac{60}{a^{4}} h_{17}
% +
% \frac{120}{a^{4}} h_{18} 
% +
% \frac{120}{a^{4}} h_{19} %\nonumber \\
% +
% \frac{120}{a^{4}} h_{20} + \cdots.
% +
% \frac{16}{a^{4}} h_{21}
% +
% \frac{16}{a^{4}} h_{22}
% +
% \frac{16}{a^{4}} h_{23} 
% +
% \frac{16}{a^{4}} h_{24}
% +
% \frac{16}{a^{4}} h_{25}.
\label{eq:Yud}
\end{align}
The terms in which the adjoint Higgs fields 
 are contracted by themselves 
 have larger pre-factors than the others.
Such terms should be practically regarded as
lower order contribution. 
For example, we should regard $h_{3}$ as 
{\it the first order} term like %as the terms of
 $h_{1}$ and $h_{2}$.
The terms of $h_{7\text{-}9}$ and $h_{14}$ 
are referred as {\it the second order} terms such as $h_{4\text{-}6}$,
and those of $h_{15\text{-}20}$ should belong to {\it the third order}
terms such as $h_{10\text{-}13}$. 

The value of 
 $\sigma$ is related to the mass of the coloured Higgs triplet
 $M_{C}$ and the GUT scale.
The GUT scale is represented as 
 $(M_{V}^{2} M_{\Sigma})^{1/3}$,
 where 
 $M_{V}$ stands for mass of 
 $X$ and $Y$ bosons ($SU(5)$ breaking gauge bosons) 
 and $M_{\Sigma}$ for the mass of $\Sigma$.
The magnitudes of these mass parameters are 
 strictly constrained  from 
 the gauge coupling unification condition\cite{MurayamaPierce,GUTscale}. 
However, the value of $\sigma$ itself can be 
 larger than the GUT scale 
 $(M_{V}^{2} M_{\Sigma})^{1/3} \simeq 2.0\times 10^{16}$ GeV by 
 taking the Higgs couplings among $H$, $\bar{H}$,
 and $\Sigma$ to be small.
Therefore, we can  
 take $a = \mathcal{O}(10\sim100)$,
 where dimension six proton decay operators
 are suppressed enough.

%\vspace{3mm}
%%%%%%%%%%%%%%%%%%%%%%%%%%%%%%%%%%%%%%%%%%%%%%%%%%%%%%%%%%%%%%%%%%%%%%
%\section{Example}

Let us now 
 illustrate a concrete example for the Yukawa
 couplings which reproduces not only 
 the realistic fermion 
 mass spectrum but also 
 the completely vanishing 
 dimension five proton decay operators.  
We take the basis where $Y_{d}$ and $Y_{e}$ are diagonal.
The coefficients of the dimension five proton decay 
 operators are 
 denoted 
\begin{align}
C_{5L}^{ijkl} \equiv Y_{ql}^{ij} Y_{qq}^{kl}, \qquad 
C_{5R}^{ijkl} \equiv Y_{ud}^{ij} Y_{eu}^{kl},
\end{align}
where $Y_{qq}$ and $Y_{eu}$ are the couplings of interactions of the
coloured Higgs field coming from the ${\bf 10}_{i} {\bf 10}_{j} H$ 
type terms.
%\footnote{%
%If we allow huge components for the matrices $f_{i}$,
%all the components of $C_{5L}^{ijkl}$ and $C_{5R}^{ijkl}$ could be 
%zero by taking $Y_{qq}$ and $Y_{eu}$ to be null matrices[8].
%However, it breaks down the reliableness of  
%the perturbative calculation, so that we here do not
%consider this situation.}.
%By requiring the realistic fermion mass spectrum and 
%vanishing the proton decay operators 
Avoiding 
 unreliable large couplings in $h_{i}$'s,
 the texture of $h_{i}$'s for the third generation 
 is uniquely 
 determined except for 
 $\Delta y_{3} \equiv y_{\tau} - y_{b}$ 
 which is 
 the difference between
 the Yukawa couplings of the tau lepton and the bottom quark.
 It should be small at the GUT scale and  
 we assume that it is 
 provided by {\it second order} terms. 
{}From Eqs.\eqref{eq:Yd}-\eqref{eq:Yud}, 
 the third generation Yukawa components 
 are induced as
\begin{align}
& (h_{6})_{33} = a^{2} \Delta y_{3}/25, \\
%%%%%
& (h_{5})_{33} 
 = a \left(  h_{2}'  \right)_{33} 
 + 2 a^{2}\Delta y_{3}/25, \\
& (h_{4})_{33} = a \left( h_{1}'  \right)_{33} 
 + a^{2} y_{b}/5 + 2 a^{2} \Delta y_{3}/25, \\
%%%%%
& (h_{3}')_{33}
 =
 -2a^{2}(y_{b}+\Delta y_{3})/75 
 -a\left( h_{1}' + h_{2}' \right)_{33}/5,
 \label{eq:h3-h0-third}
\end{align}
up to the {\it second order}. 
%where $h_{1,2}' \equiv h_{1,2} + 30 h_{7,8}/a^{2}$ and
Here, $h_{3}' \equiv h_{3} - h_{9}/a + 30 h_{14}/a^{2}$.
In order to avoid 
 $\mathcal{O}(a^{2} y_{b})$ terms
 in $(h_{5})_{33}$ and
 $(h_{4})_{33}$,
 we must take 
\begin{align}
 (h_{1})_{33} = -a y_{b}/5, \qquad (h_{2})_{33} = 0. 
\end{align} 
Then, the value of $(h_{3})_{33}$ is determined 
 up to order $\mathcal{O}(a^{2} y_{b})$
 as 
\begin{align}
 (h_{3})_{33} = a^{2} y_{b}/75. 
\end{align}
This term should be regarded as the 
 {\it first order} term % $1/a$
 because of the large pre-factor $1/75$,  
% (coming from a self contract of $\Sigma$'s $SU(5)$ index)
% cancels one $a$ factor,
 in which the value of $(h_{3})_{33}$ itself 
 is kept of $\mathcal{O}(1)$. 
Now all the third generation components of $h_{i}$'s 
are determined. 
The couplings for the second generation are also 
 determined in a similar way. 
Those for the first generation
 are not uniquely determined since 
 there are large degrees of freedom 
 in the {\it third order} terms.

Summarizing above discussions, 
 the {\it first order} terms are
 uniquely determined as 
\begin{gather}
%%%%%
 h_{1}
 =
 -\frac{a}{5}\text{diag}
 \left(
 0, 
 0,
 y_{b} 
 \right), 
\quad
%%%%%
 h_{2}
 = {\bf 0}, \nonumber \\
%%%%%
 h_{3}
 =
 \frac{a^{2}}{75} \text{diag}
 \left(
 0,
 0, 
 y_{b} 
 \right).
%%%%%
\label{eq:first}
\end{gather}
The {\it second order} terms are
 also determined almost automatically as 
\begin{align}
 h_{4}
 =&
 \frac{a^{2}}{75}
 \text{diag}
 \left(
 0,
 9y_{s}+y_{\mu}, 
 \Delta y_{3} 
 \right),\nonumber  \\
%%%%%
 h_{5}
 =&
 \frac{a^{2}}{75}
 \text{diag}
 \left(
 0,
 - 6y_{s}+y_{\mu}, 
 \Delta y_{3} 
 \right), \nonumber \\
%%%%%
 h_{6}
 =&
 \frac{a^{2}}{25}
 \text{diag}
 \left(
 0,
 -y_{s}+y_{\mu}, 
 \Delta y_{3} 
 \right),   \nonumber \\
%%%%%
 h_{7}= h_{8}
 =&
 -\frac{a^{3}}{450}
 \text{diag}
 \left(0,  y_{\mu},  \Delta y_{3}\right), \nonumber \\
%%%%%
 h_{9} =& h_{14} = {\bf 0}.
\label{eq:second}
\end{align}
For the {\it third order} terms,
 there are various choices, and
 one example is 
\begin{align}
%%%%%
 h_{10} =&h_{16\text{-}25} = {\bf 0}, \nonumber \\
%%%%%
 h_{11} =&
 \frac{a^{3}}{25}
 \text{diag}
 \left(
 -y_{d} + 2 y_{e}/7, 0 ,0
 \right), \nonumber \\
%%%%%
 h_{12} =&
 \frac{a^{3}}{25}
 \text{diag}
 \left(
  3y_{d}/2 -  2y_{e}/3, 0, 0
 \right),\nonumber \\
%%%%%
 h_{13} =&
 \frac{a^{3}}{25}
 \text{diag}
 \left(
 - y_{d}/2 -  y_{e}/3, 0 ,0
 \right),\nonumber \\
%%%%%
 h_{15} =& 
 \frac{4a^{4}}{3675} \text{diag}
 \left(y_{e},0,0\right).
\label{eq:third}
\end{align}
It is worth noting that 
 the accurate Yukawa couplings of the 
 down-sector quarks and charged leptons 
 are obtained as 
\begin{align}
Y_{d} = \text{diag}(y_{d}, y_{s}, y_{b}), \qquad 
Y_{e} = \text{diag}(y_{e}, y_{\mu}, y_{\tau}), 
\label{eq:Yd-Ye-result}
\end{align}
as well as the couplings of coloured Higgs triplet 
 vanish as  
\begin{align}
Y_{ql} = Y_{ud} = {\bf 0}.
\label{eq:YqlYudeq0}
\end{align}
Once these Yukawa interactions are realized at the GUT scale, 
 any dimension five proton decay process will not appear
 even if the renormalization 
 group equation (RGE) effects are taken into account.

%%%%%%%%%%%%%%%%%%%%%%%%%%%%%%%%%%%%%%%%%%%%%%%%%%
\begin{table}[thb]
\begin{tabular}{c|c|c|c|c}
\hline \hline 
       & {\it order}  &        & Example 1        & Example 2 \\
\hline 
$W_{1}$& 1st &$h_{1}$ & \multicolumn{2}{c}{$(0,0,-0.64)$} \\
       &     &$h_{2}$ & \multicolumn{2}{c}{{\bf 0}} \\
\hline
$W_{2}$& 1st &$h_{3}$ & \multicolumn{2}{c}{$(0,0,2.3)$} \\ \cline{2-5}
       & 2nd &$h_{4}$ & \multicolumn{2}{c}{$(0,0.72,0.40)$} \\
       &     &$h_{5}$ & \multicolumn{2}{c}{$(0,-0.21,0.40)$} \\
       &     &$h_{6}$ & \multicolumn{2}{c}{$(0,0.29,1.2)$} \\
\hline
$W_{3}$& 2nd &$h_{7}$  &  \multicolumn{2}{c}{$(0,-1.5,-3.6)$}\\
       &     &$h_{8}$  &  \multicolumn{2}{c}{$(0,-1.5,-3.6)$}\\ 
       &     &$h_{9}$  &  \multicolumn{2}{c}{{\bf 0}}\\ \cline{2-5}
       & 3rd &$h_{10}$ &  {\bf 0}        & {\bf 0} \\ 
       &     &$h_{11}$ & $(-0.48,0,0)$   & {\bf 0}\\
       &     &$h_{12}$ & $(0.68,0,0)$    & {\bf 0}\\ 
       &     &$h_{13}$ & $(-0.29,0,0)$   & {\bf 0}\\
\hline
$W_{4}$& 2nd &$h_{14}$ &  \multicolumn{2}{c}{{\bf 0}}\\ \cline{2-5}
       & 3rd &$h_{15}$ & $(0.19,0,0)$      & $(-0.65,0,0)$\\ 
       &     &$h_{16}$ &  {\bf 0}          &{\bf 0} \\
       &     &$h_{17}$ &  {\bf 0}          & {\bf 0}\\
       &     &$h_{18}$ & {\bf 0}           & $(3.3,0,0)$\\
       &     &$h_{19}$ & {\bf 0}           & $(-1.4,0,0)$\\
       &     &$h_{20}$ & {\bf 0}           & $(-0.71,0,0)$\\ \cline{2-5}
%       & 4th &$h_{21\text{-}25}$ &  \multicolumn{2}{c}{{\bf 0}}\\
\hline \hline
\end{tabular}
\caption{Examples of the matrix elements of $h_{i}$'s at the GUT scale 
 which reproduce the realistic fermion mass spectrum and vanish the
 proton decay operators. 
 Here, we take $a=55$ and $\tan\beta=10$.
 We take the diagonal basis of the down-type quark and 
 the charged lepton mass matrices. 
 The flavour mixing is imposed into 
 the up-sector Yukawa couplings.}
\label{Tab:texture}
\end{table}
In Table I, 
 we present
 the magnitudes of Yukawa couplings at the
 GUT scale by using the results in Ref.\cite{KoideFusaoka}.
The applicability of the perturbation 
 (couplings$\lesssim \sqrt{4\pi}$) 
 is satisfied for all components.
%It is also satisfied  
% even in  rather large $\tan\beta$ ($\tan\beta=10$) region.
Notice that it can be satisfied even  
 in  
 the rather large $\tan\beta$ ($\tan\beta\sim 10$) region. 
Ordinal $SU(5)$ GUT models, e.g. with decoupling SUSY 
 breaking spectrum,
 should have a small $\tan\beta$
 of order 1\cite{consistentmodel}.

The Yukawa couplings of the up-type quarks can 
 also be derived from 
 higher dimensional terms.
We assume that the Yukawa coupling of the top quark comes from $Y_{1}$ in
 Eq.\eqref{2}, that of the charm quark from $W_{2}$, and 
 that of the up quark from $W_{4}$.
For example, supposing the simple superpotential, 
\begin{align}
W = \frac{\epsilon_{abcde}}{4}
 {\bf 10}^{ab}_{i} {\bf 10}^{cd}_{j} H^{e}
 \left\{
 f^{ij}_{\rm c} %{\bf 10}^{ab}_{i} {\bf 10}^{cd}_{j} H^{e}
 \frac{(\Sigma \Sigma)}{M_{\rm Pl}^{2}} 
 + 
% \frac{\epsilon_{abcde}}{4}
 f^{ij}_{\rm u} %{\bf 10}^{ab}_{i} {\bf 10}^{cd}_{j} H^{e}
 \frac{(\Sigma \Sigma \Sigma \Sigma)}{M_{\rm Pl}^{4}}
 \right\},
 \nonumber 
\end{align}  
we obtain the Yukawa matrices
\begin{align}
Y_{u} = Y_{1} + \frac{30}{a^{2}}f_{\rm c} + \frac{210}{a^{4}} f_{\rm u}.
\nonumber 
\end{align}
In the basis 
 $Y_{u}$ should be given by 
 $Y_{u} = U_{\rm CKM}^{\dagger} Y_{u}^{\rm diag}$. 
% the matrices $f$'s is also the origin of the 
% generation mixing. % in this scenario.
This example gives 
\begin{align}
(Y_{1})_{33} = 0.75, \quad (f_{\rm c})_{22} = 0.18, \quad 
(f_{\rm u})_{11}= 0.26, \nonumber 
\end{align}
for $Y_u^{\rm diag}$ with
the same values of $a$ and $\tan\beta$ 
in Tab.\ref{Tab:texture}.

%\vspace{3mm}
%%%%%%%%%%%%%%%%%%%%%%%%%%%%%%%%%%%%%%%%%%%%%%%%%%%%%%%%%%%%%%%%%%%%%%
%\section{Discussion and summary}

Some comments are in order. 
The first is about 
 the coefficients of the proton decay operators. 
They 
 must include at least one first generation 
 quark superfield.
Therefore, it is not necessary to eliminate 
 all components of the coloured Higgs Yukawa couplings as 
 in Eq.\eqref{eq:YqlYudeq0}. 
In fact we can realize such Yukawa matrices 
 which have more choices than 
 the examples shown above. 
However, in this case 
 the RGE effect must be taken into account to estimate 
 the proton decay rate, 
 since the second and third generation 
 components of $Y_{ql}$ and $Y_{ud}$ are
 transmitted into 
 the first generation components 
 through the generation mixings.
The RGE analysis shows that  
 the large entry of the second and third generation components
 could be destructive\footnote{%
In Ref.[8], 
 the authors adopted the ingenious texture which they referred as 
 {\it the consistent model 1}.
This model has the non-zero first generation components,
 however, they avoided the large RGE effects 
 by vanishing second and third generations' components in 
 $Y_{ql}$. 
This scenario is effective only in the case of 
 small $\tan\beta$ such as $\tan\beta ={\cal O}(1)$. }.
This means 
 that RGE effects will break 
 the proton stability
 even if 
 all the first generation components of $Y_{ql}$ and $Y_{ud}$ are 
 zero. 
If $Y_{ql}$ and $Y_{ud}$ do not include 
 the first generation components
 at the nucleon mass scale (not 
 at the GUT scale), 
 the dimension five proton decay
 processes will disappear as pointed out in Ref.\cite{examples}. 
It is an interesting possibility. 
However, 
 there must be a reason why the scale 
 is not the GUT scale but the nucleon 
 mass scale.

The second comment is about the contribution %to the proton decay processes 
 from the sub-leading effects.
When the soft SUSY breaking tri-linear scalar interactions
 of the coloured Higgs 
\begin{align}
 -\mathcal{L}_{\rm soft}
 \supset
 A_{ql}^{ij} \tilde{Q}_{i} \epsilon \tilde{L}_{j} \bar{H}_{C}
 +
 A_{ud}^{ij} \tilde{u}_{Ri}^{*} \tilde{d}_{Rj}^{*} \bar{H}_{C} 
 +
 {\rm H.c.},
\end{align}
are introduced, $A_{ql} Y_{qq}$ and $A_{ud} Y_{eu}$ 
 can contribute to the proton decay processes. 
We have neglected 
 these $A$-term contributions 
 in the above discussions, which 
 can be justified 
 in the minimal supergravity context.
It is because 
 these terms are generally 
 proportional to the corresponding Yukawa couplings. 
Therefore, when they do not exist at the GUT scale, 
 there will be no contribution 
 at the low energy scale even if we   
 take account into the RGE effects.

We have tried to reproduce the suitable 
 fermion mass hierarchy as well as 
 to suppress the proton decay 
 in the $SU(5)$ SUSY GUT framework 
 with the minimal field contents.
The realistic fermion mass spectrum can be realized 
 simultaneously with 
 vanishing dimension five proton decay processes.
These requirements have almost determined 
 the Yukawa couplings of 
 higher dimensional operators at the GUT scale.
%Our model would be reasonable 
%since the theory at the GUT scale should be represented as 
%the effective theory of the fundamental one, 
%and it should include the higher dimensional terms.

%%%%%%%%%%%%%%%%%%%%%%%%%%%%%%%%%%%%%%%%%%%%%%%%%%%%%%%%%%%%%%%%%%%%%%
%\acknowledgments

%\vspace{3mm}

The authors would like to thank N. Okada and M. Kakizaki 
for useful discussions 
and be grateful to 
 A. Merle for reading the manuscript carefully.
N.H. is supported by Alexander von 
 Humboldt Foundation.

%%%%%%%%%%%%%%%%%%%%%%%%%%%%

\end{document}